\documentclass[doublecol]{epl2}
\usepackage{amssymb}
\usepackage{amsmath}

\newcommand{\eqb}{\begin{eqnarray}}
\newcommand{\eqe}{\end{eqnarray}}

\def\bes{\begin{subequations}}
\def\ees{\end{subequations}}

\title{Hysteresis Loops and Multi-stability: From Periodic Orbits to Chaotic
Dynamics (and Back) in  Diatomic Granular Crystals}
\shorttitle{Hysteresis and Multi-stability in Diatomic Granular Crystals}

\author{C. Hoogeboom\inst{1} \and Y. Man\inst{2} \and N. Boechler\inst{3}
\and G. Theocharis\inst{2} \and P.G. Kevrekidis\inst{1} \and
I.G. Kevrekidis\inst{4} \and  C. Daraio\inst{5}}
\shortauthor{C. Hoogeboom \etal}

\institute{
\inst{1} Department of Mathematics and Statistics, University of Massachusetts, Amherst, MA 01003-4515, USA \\
\inst{2} Graduate Aerospace Laboratories (GALCIT) California Institute of Technology, Pasadena, CA 91125, USA \\
\inst{3} Department of Mechanical Engineering, Massachusetts Institute of Technology, Cambridge, MA 02139, USA \\
\inst{4} Department of Chemical and Biological Engineering
and PACM, Princeton University, Princeton, NJ, 08544, USA
}

\pacs{45.70.-n}{Granular systems}
\pacs{63.20.Pw}{Localized modes}
\pacs{05.45.-a}{Nonlinear dynamics and chaos}

\abstract{In the present work we consider a diatomic granular crystal, consisting of alternating aluminum and steel spheres, where the first sphere is an 
aluminum one. 
	The combination of dissipation, driving of the boundary, and intrinsic nonlinearity leads to complex dynamics. 
	Specifically, we find that the interplay of nonlinear surface modes with modes created by the driver create the possibility, as the driving amplitude is increased, of limit cycle saddle-node bifurcations beyond which the dynamics of the system becomes chaotic. 
	In this chaotic state, part of the applied energy can propagate through the chain.
	We also find  that the chaotic branch depends weakly on the driving frequency and speculate a connection between the chaotic dynamics with the gap openings between the spheres. 
Finally, a reverse parametric continuation reveals hysteretic dynamics and the existence of an interval of multi-stability 
involving stable periodic solutions and genuinely chaotic ones. 
	The computational identification and theoretical interpretation of the bifurcation diagram of the system is corroborated by direct experimental measurements showcasing all of the above features.
}

\begin{document}

\maketitle

{\it Introduction.}\label{S:______Introduction}
	 Granular systems, consisting of densely-packed particles that interact through nonlinear, tensionless potentials, have been largely recognized in the last decade as a fertile testbed where ideas from nonlinear dynamics can be 
put to good use~\cite{nesterenko1,sen08}.
	A significant advantage of these systems is their tunability, which enables the access of nearly linear, weakly nonlinear, and highly nonlinear regimes. 
	Relevant investigations have predominantly focused on the dynamics of nonlinear waveforms, including traveling waves~\cite{nesterenko1,nesterenko0,sen08,coste97,our08,s1,s2,g1} and discrete breathers \cite{our10}, as well as other nonlinear processes, including second harmonic generation and nonlinear resonances \cite{VT}. 
	This direction of research has already led to numerous potential applications, including energy absorbing layers \cite{dar06,hong05,doney06}, sound scramblers \cite{dar05b}, acoustic lenses \cite{Spadoni}, and rectifiers \cite{Diode}.  
	
In the present work, we explore another important dimension of granular systems, which involves their ability to support chaotic dynamics accompanied by energy transmission when subjected to external driving  
above a critical amplitude in the forbidden frequency gap. 
	Intrinsic dissipation is abundant in the system and has been the subject of intense recent investigation~\cite{nesterenko2,us09,vergara}.
	Combining the ability to dynamically drive the system~\cite{Diode} with this intrinsic dissipation, we produce a case example of a damped driven system of coupled nonlinear oscillators. This system has the potential for the future study of pattern formation~\cite{cross,khomeriki2001} as well as nonlinear supratransmission \cite{geniet2002,khomeriki2004}, triggered by the 
interplay of ordered (periodic, quasi-periodic) and 
chaotic dynamics~\cite{kopidakis}.

	Our starting point will be a detailed analysis of the periodic modes of the damped-driven diatomic system, consisting of alternating aluminum and steel spheres, where the first sphere is aluminum. 
	The interesting feature here is that the presence of harmonic driving at the boundary with a frequency within the band gap of the underlying linear
spectrum and its interplay with dissipation, nonlinearity, and discreteness enables two classes of relevant states. These are nonlinear surface modes as well as ones tuned to the external actuator.
	These two branches of time-periodic 
solutions are observed to collide and disappear in a limit cycle
saddle-node bifurcation.
	Beyond this critical point no stable, periodic solutions are found to exist and the dynamics is found to ``jump'' to a chaotic branch (as the strength of the drive is further increased). 
	This jump is accompanied by the onset of energy transmission through the chain.
	On the other hand, as the drive is decreased along the chaotic branch, a hysteretic loop is found to arise, with the chaotic solutions persisting well below this periodic saddle-node point. 
	Such a hysteretic loop between periodic solutions and chaotic ones can be found in other physical systems ranging from nonlinear optics (e.g. dissipative soliton molecules in mode-locked fiber lasers~\cite{zavyalov}) to droplet dynamics (e.g. in dripping faucets~\cite{jpsj}), yet it has rarely been experimentally explored. 

\begin{figure}
\includegraphics[width=.45\textwidth]{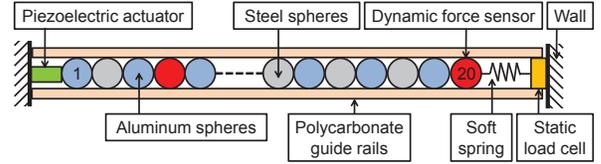}
\caption{Schematic of the experimental setup.}
\label{f:experimental_setup}
\end{figure}

{\it Experimental Setup.}\label{S:______Experimental_setup}
	We construct a diatomic granular crystal, as shown in Fig.~\ref{f:experimental_setup},  by alternating $N = 20$ aluminum spheres (6061-T6 type, radius $R_{a} = 9.53$ mm, mass $m_{a} = 9.75$ g, elastic modulus $E_{a} = 73.5$ GPa, Poisson ratio $\nu = 0.33$) and stainless steel spheres (316 type, $R_{b} = R_{a}$, $m_b = 28.84$ g, $E_b = 193$ GPa, $\nu = 0.3$)~\cite{OurDimer21,OurDefects}. 
	We constrain the spheres to a one-dimensional configuration by using four polycarbonate rods arranged in a square pattern, which are aligned by periodically spaced guide plates. 
	On a steel block fixed to the table at one end of the granular crystal, we mount a piezoelectric actuator that applies dynamic displacements to the crystal boundary. 
	We calibrate the displacement of the actuator tip as a function of driving voltage by using a laser vibrometer in conjunction with a strain-gauge embedded in the actuator. 
	For our granular crystal, the particle next to the actuator is an aluminum particle. 
	At the opposite end of the crystal from the actuator, we apply a static load ($F_0 = 8$ N) by compressing a soft spring ($K_s = 1.24$ kN/m) between the last particle of the crystal, and a static load cell mounted in a teflon holder. 
	The spring, static load cell, and granular crystal are compressed by fixing a second steel block to the table at some distance relative to the one at the other end. 
	The static load is measured directly by the in-line static load cell. 
	We measure the time variation of propagating stress waves caused by the driver by placing calibrated piezoelectric disks inside two halves of a steel sphere (maintaining properties such as inertia and bulk stiffness of the original steel spheres) \cite{dar06}. 
	For our frequencies of interest, each sensor measures the average force between each of its two adjacent contacts. 
	We place sensor particles at sites $n = 4$ and $n = 20$. 
	The measured dynamic force signals are then filtered using a low pass filter and amplifier (Alligator Tech, 30kHz 8th order Butterworth low-pass filter) and recorded with a data acquisition board. 
	The acquired signals are then digitally filtered with a 100 Hz 5th order Butterworth high pass filter to remove 60 Hz electrical noise. 

{\it Model.}\label{S:______Model}
	We model our damped and driven (through actuating the boundary) granular crystal, which consists of alternating steel and aluminum beads under a precompressive force $F_0$, as the following set of equations:

\begin{align} 
m_j\ddot{u}_j =& A_{j-1,j}[\delta_{j-1,j}-(u_j-u_{j-1})]_+^{3/2} 
\nonumber
\\
&-
A_{j,j+1}[\delta_{j,j+1}-(u_{j+1}-u_j)]_+^{3/2} 
\nonumber
\\
&-
\frac{m_j\dot{u}_j}{\tau} 
\label{e:motion}
\end{align}
where $m_j$ is the mass of the $j$th bead, $A_{j,j+1}$ is the coefficient of elasticity between beads $j$ and $j+1$~\cite{nesterenko1},
and $\delta_{j,j+1}=(F_0/A_{j,j+1})^{2/3}$ is the amount of static overlap of the beads $j$ and $j+1$ (due to the precompression) under equilibrium conditions. 
	The actuator is placed at the left edge of the chain and provides a prescribed displacement with a frequency $f_d$ and amplitude $\alpha$ given by $u_0(t)=\alpha\cos(2\pi f_d t)$; the right edge of the chain is free.
	The subscript  accounts for the fact that if two beads are not in contact, they do not exert any force on each other (zero-tension).
	The dissipation is chosen to be a uniform dashpot form throughout the chain, $\tau=1.75$ ms in line with the recent exposition of~\cite{Diode}.
	The linear spectrum of the undamped, undriven chain has an acoustic band and 
an optical band, with a finite gap between them, and a semi-infinite gap above 
the optical band \cite{our10}. 
	For a precompressive force of $F_0=8$ N, using the standard values of
material parameters, the theory predicts that the acoustic band runs from 0 kHz to 3.98 kHz, while the optical band runs from 6.96 kHz to 8.02 kHz. 

\begin{figure}[h!]
\includegraphics[width=.4\textwidth]{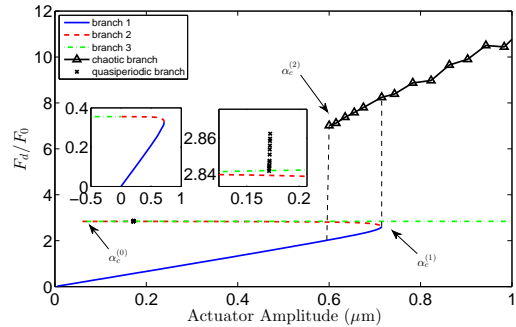}
\caption{
	Bifurcation diagram for the continuation in the actuator amplitude $\alpha$ as a bifurcation parameter, illustrating the average dynamical force amplitude, normalized against the static load $F_0$, at the fourth site of the chain. 
	There are three branches of periodic nonlinear modes (stable solid blue, strongly unstable dashed red, and -oscillatorily- unstable dash-dotted green). 
Limit cycle saddle-node bifurcations arise for $\alpha= \alpha_c^{(0)}$ between the dashed and dash-dotted branches and for $\alpha= \alpha_c^{(1)}$ between the solid and dashed. 
	Past $\alpha=\alpha_c^{(2)}$, an additional large amplitude chaotic dynamical branch also exists.
For completness we also note the existence of a Hopf bifurcation in branch 3	which spawns a branch of quasiperiodic solutions, shown as `x' 
marks in the figure.
	An enlarged view of the Hopf bifurcation is shown in the right insert.
	The left insert shows the corresponding 
bifurcation diagram for the undamped system, in which branch 
2 reaches $\alpha=0$ and extends towards branch 3
for $\alpha<0$.}
\label{f:bifurcation_diagram}
\end{figure}

{\it Numerical Results and Analysis.}\label{S:______Numerical_results}
	Our main theoretical findings are summarized in the bifurcation diagram of Fig. \ref{f:bifurcation_diagram}, which has been computed for $f_d = 6$ kHz. 
	We have sought periodic solutions to the system, using the amplitude of the actuator  as a bifurcation parameter (for fixed frequency) in our Newton algorithm for the damped-driven dynamical system of Eq.~(\ref{e:motion}). 
	Other examples of the realization of such a computational method for
granular crystals can be found e.g. in~\cite{our10,Diode} (see also references therein).  

	The bifurcation diagram showcases three principal branches of periodic solutions that are present in our diatomic chain of 20 spheres, for which the particle next to the actuator is an aluminum sphere. 
	The branch represented by the (blue) solid line corresponds to modes that are  absent when the actuator is off. 
	On the contrary, the second (red) dashed and third (green) dash-dotted line appear to stem from a finite force, well above the precompression threshold emerging through a limit cycle saddle-node bifurcation at $\alpha_c^{(0)} \approx 6 \times 10^{-8}$ m. 
	In fact, in the variant of the model without the dissipation these two branches merge at $\alpha=0$ m, as shown in the left inset of Fig. \ref{f:bifurcation_diagram}. 
	Both naturally emerge as a driven continuation of nonlinear modes, which are the surface analogues of intrinsic bulk discrete breathers that the dimer chain has been shown to sustain \cite{our10}. 
	The finite (small $\alpha_c^{(0)}$) critical threshold for their existence is purely a byproduct of the inability of an undriven chain to sustain large amplitude localized modes in the presence of finite damping. 
	
\begin{figure}[h!]
\begin{center}
\includegraphics[width=.2\textwidth]{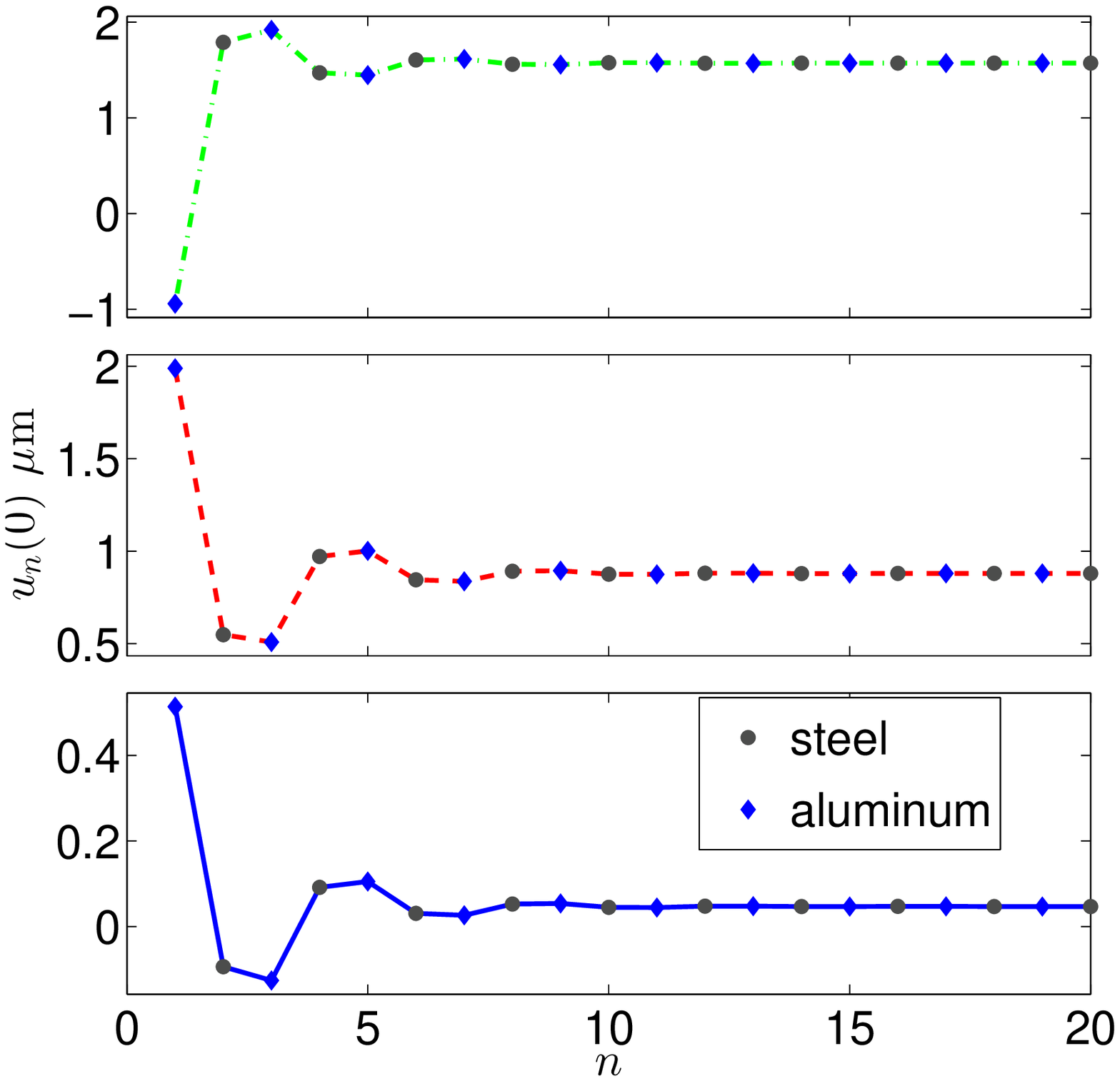}
\includegraphics[width=.2\textwidth]{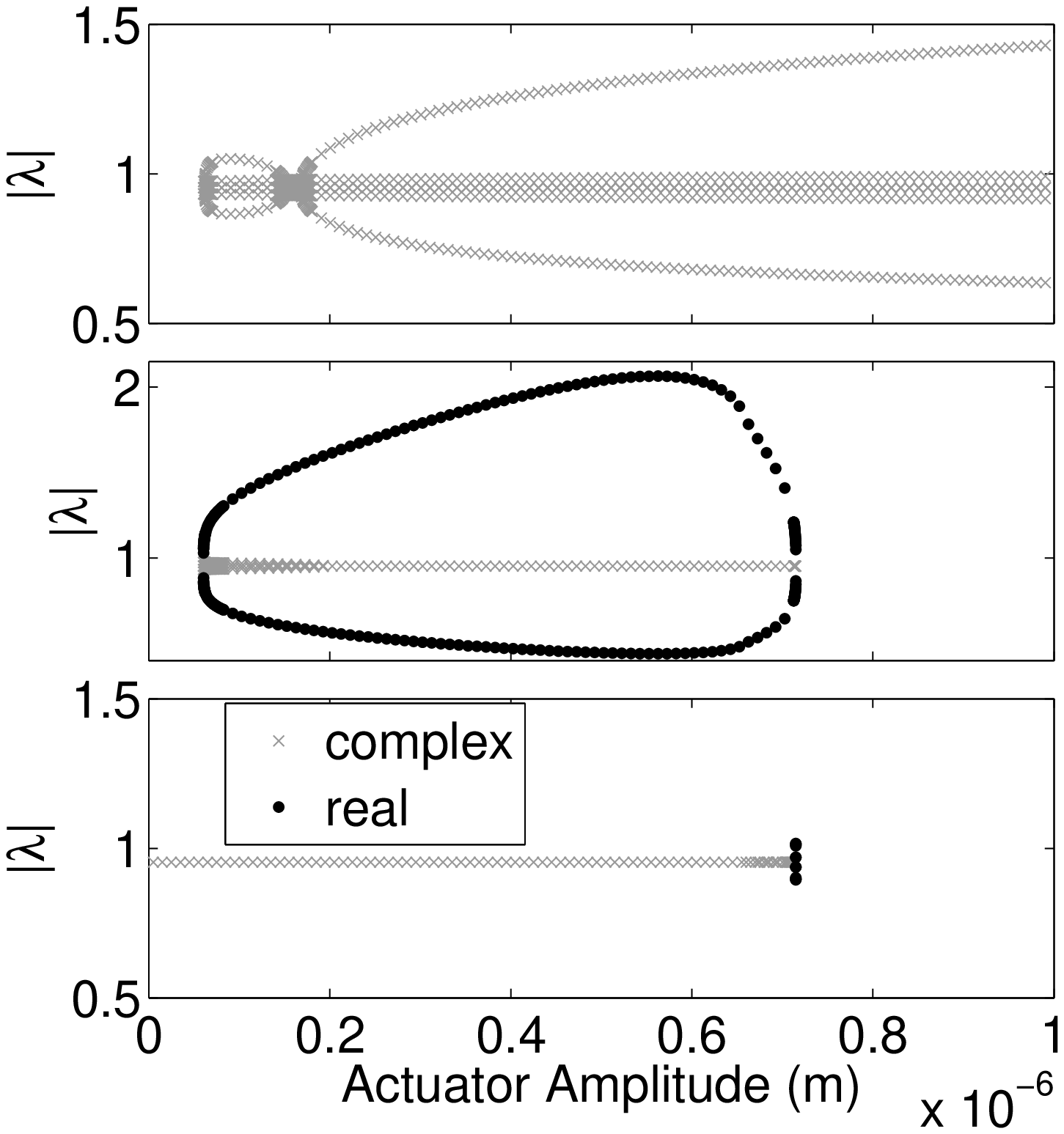}
\end{center}
\caption{
	Sample profiles (left panels) from each of the three breather branches are shown (at $t=0$) for $\alpha=4 \times 10^{-7}$ m indicating the progression (from bottom to top) from lower amplitudes to higher ones, and from in-phase (with $u_1(0)>0$) to out-of-phase (with $u_1(0)<0$) structure with respect to the actuator (with $u_0(0)>0$).
	The lines connecting the lattice sites in each figure are colored the same as the corresponding branch in \ref{f:bifurcation_diagram}.
	The continuation of the absolute value of the Floquet multipliers of these branches (right panels) illustrates the absence of instability for the bottom branch, the strong instability (due to a real -black symbol- pair of non-unit magnitude multipliers) of the middle one and the oscillatory instability (due to a complex -gray symbol- pair)  of the top one.} 
\label{f:damped_floquet}
\end{figure}

A typical profile of a nonlinear localized mode from each of the three branches at $\alpha=4\times 10^{-7}$ m, and the stability properties of the entire branches (through the modulus of the Floquet multipliers associated with such periodic orbits) are given in Fig.~\ref{f:damped_floquet}. 
	These nonlinear surface modes are very close to the profiles of the hybrid bulk-surface localized solutions (see Fig.~9 in~\cite{our10}). 
	Furthermore, the solid and dashed branches shown at $t=0$ are essentially in phase with the driver (but of lower and higher amplitude, respectively), while the dash-dotted one bears a (large amplitude) principally out-of-phase response with the actuator.
	Recall that the driver is cosinusoidal, and bears a positive amplitude at $t=0$.
	Accordingly, we draw the above conclusions on the basis of the position of the (first) dominant amplitude site of the corresponding surface localized modes.
	While the low-amplitude response to the driver is an asymptotically stable one, as illustrated through the absence of real (or complex) Floquet multipliers in the corresponding stability analysis of Fig. ~\ref{f:damped_floquet}, the unstable dashed branch always bears at least a real pair of multipliers which has $|\lambda|>1$, and the (oscillatorily) unstable dash-dotted branch bears quartets of complex multipliers (in the exception of two very small intervals of $\alpha$'s (in a neighborhood of the initial bifurcation point and of 
$2 \times 10^{-7}$ m).

	The key feature of interest within the system is then that the low amplitude periodic solution branch induced by the actuator and the unstable dashed branch due to the system's intrinsic linear and nonlinear properties~\cite{our10} collide and {\it disappear} for $\alpha_c^{(1)} \approx 7.14 \times 10^{-7}$ m. 
	What we observe past this critical response indicates a chaotic large amplitude response of the system, which appears to generically exist for amplitudes $\alpha > \alpha_c^{(1)}$. 

	In order to identify the domain of existence of this chaotic branch, we have also tried to continue this branch for $\alpha < \alpha_c^{(1)}$. 
	Indeed, we have been able to identify such solutions down to $\alpha_c^{(2)} \approx 5.9 \times 10^{-7}$ m. 
	It should be stressed that these ``solutions'' are obtained via numerical integration of the ordinary differential equations of motion only and are not numerically exact solutions.
	Taking an initial condition ``close'' to the chaotic attractor, 
we then integrate it forward in time. 
	If the solution stays close to the chaotic attractor after 4000 oscillations of the actuator, then we designate this solution as a part of the chaotic branch. 
	For values of $\alpha < \alpha_c^{(2)}$ even large amplitude initial data have been found to asymptotically tend to the low amplitude periodic solution of the system (prior to the 4000 oscillation cutoff). 
	Hence, this last apparent critical point signals the boundary of existence (or at least of dynamical stability/attractivity within this
reporting horizon) of the relevant chaotic branch. 
	This, in turn, implies that between these two critical points, i.e., for $\alpha_c^{(2)} < \alpha < \alpha_c^{(1)}$, between the emergence of the stable chaotic branch and the termination of the asymptotically stable low amplitude periodic solution branch induced by the driver, the dynamics of the system will be multi-stable. 
	Namely, for low initial condition amplitudes, the dynamics will converge to the small amplitude periodic solutions, while for large such amplitudes/forces, it will be attracted to the chaotic orbits shown in Fig.~\ref{f:bifurcation_diagram}. 
	For $\alpha > \alpha_c^{(2)}$ and suitable initial data, positive Lyapunov exponents have been found to corroborate the chaotic nature of the relevant orbits. 
	Interestingly, it appears that the role of the separatrix between low amplitude ordered dynamics and large amplitude chaotic ones in this multi-stable regime involves the metastable green dash-dotted breather branch. 
	It is conceivable that the stable and (low-dimensional) unstable manifolds of this branch are involved in the ``demise" of the chaotic attractor observed  in our dynamical simulations.

{\it Experimental Results.}\label{S:______Experimental_results}
	We first characterize the near-linear response of our $N = 20$ particle diatomic granular crystal, as described in Ref. \cite{our10}. 
	We experimentally measure the linear spectrum of the diatomic chain by applying low-amplitude (approximately $5$ mN peak), broadband frequency ($2-18$ kHz), uniform noise for $1040$ ms with the piezoelectric actuator. 
	Using the force-time signal measured at each sensor location, we then calculate the power spectral density (PSD) of the central portion of the force response ($524.3$ ms duration), and divide this by the PSD of the voltage signal (for the same time window) applied to the actuator. 
	We then average this PSD ratio over 8 trials. 
	We obtain our transfer function by normalizing this averaged PSD ratio by the mean PSD level between $2-4$ kHz (which corresponds to the center of the acoustic band). 
	The cutoff frequencies are then determined by finding frequencies where the attenuation is greater than $-10$ dB. 
	In our experiments, the linear spectrum was measured before and after the bifurcation experiments in order to characterize any change in the cutoff frequencies due to the dynamic loading of the crystal. 

	The linear spectrum measured initially (before any single frequency experiments) exhibits an acoustic band cutoff frequency of $4.37$ kHz, a lower optical band cutoff frequency of $7.83$ kHz, and an upper optical band cutoff frequency of $8.50$ kHz. Namely, all the cutoff frequencies show a systematic upshift of about 9\% compared to the theoretical predictions (see above). Possible explanations can be found in \cite{our10}, where a similar upshift has been observed.

However, a further upshift has been observed in the spectrum measured following the maximum amplitude driving (after $1.2$ $\mu$m driving amplitude with $f_d=6$ kHz), which has cutoff frequencies of $4.83$, $8.54$, and $9.45$ kHz, respectively. 
	We speculate that changes in the linear spectrum throughout the course of the experiment, could be caused by slight dynamic rearrangement of the granular crystal under high loading conditions. 
	Under such conditions, particles are expected to lose contact, and the crystal may 
reassemble into an arrangement with slightly modified off axis displacements. 

\begin{figure}[h!]
\includegraphics[width=.4\textwidth]{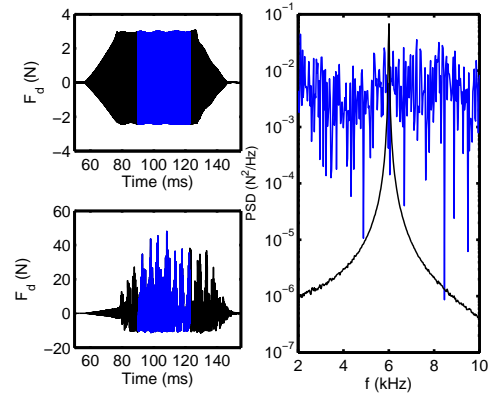}
\caption{\label{f:force_time} 
	(Left) Experimental force-time histories measured by the sensor at $n = 4$. 
	The blue is the region over which the PSD is calculated. 
	The top left figure corresponds to the largest low amplitude periodic response (blue square in Fig \ref{f:bifurc_diagram}), while the bottom left figure corresponds to the first amplitude following the onset of chaos (blue diamond in Fig \ref{f:bifurc_diagram}). 
	(Right) PSD corresponding to the top left 
(black) 
and bottom left 
(blue) 
force-time histories.}
\end{figure}

	Following the characterization of the linear spectrum, we experimentally characterize the bifurcation of the system response by exciting the crystal with a single frequency signal  ($6$, $6.5$, and $6.8$ kHz) applied via the piezoelectric actuator. 
	Each signal is $90$ ms in duration (where the amplitude $\alpha$ of the signal is linearly increased and decreased during first and last $20$ ms, respectively). 
	This linear ramp enables us to track 
the low amplitude branch of stable periodic solutions. 
	For a given driving frequency, we sequentially increase the maximum amplitude of the driving signal $\alpha$, and record the force measured by the sensor at particle $n = 4$. 
	Examples of the measured force-time response can be seen in the left panels of Fig. \ref{f:force_time}. 
 
\begin{figure}[h!]
\includegraphics[width=.225\textwidth]{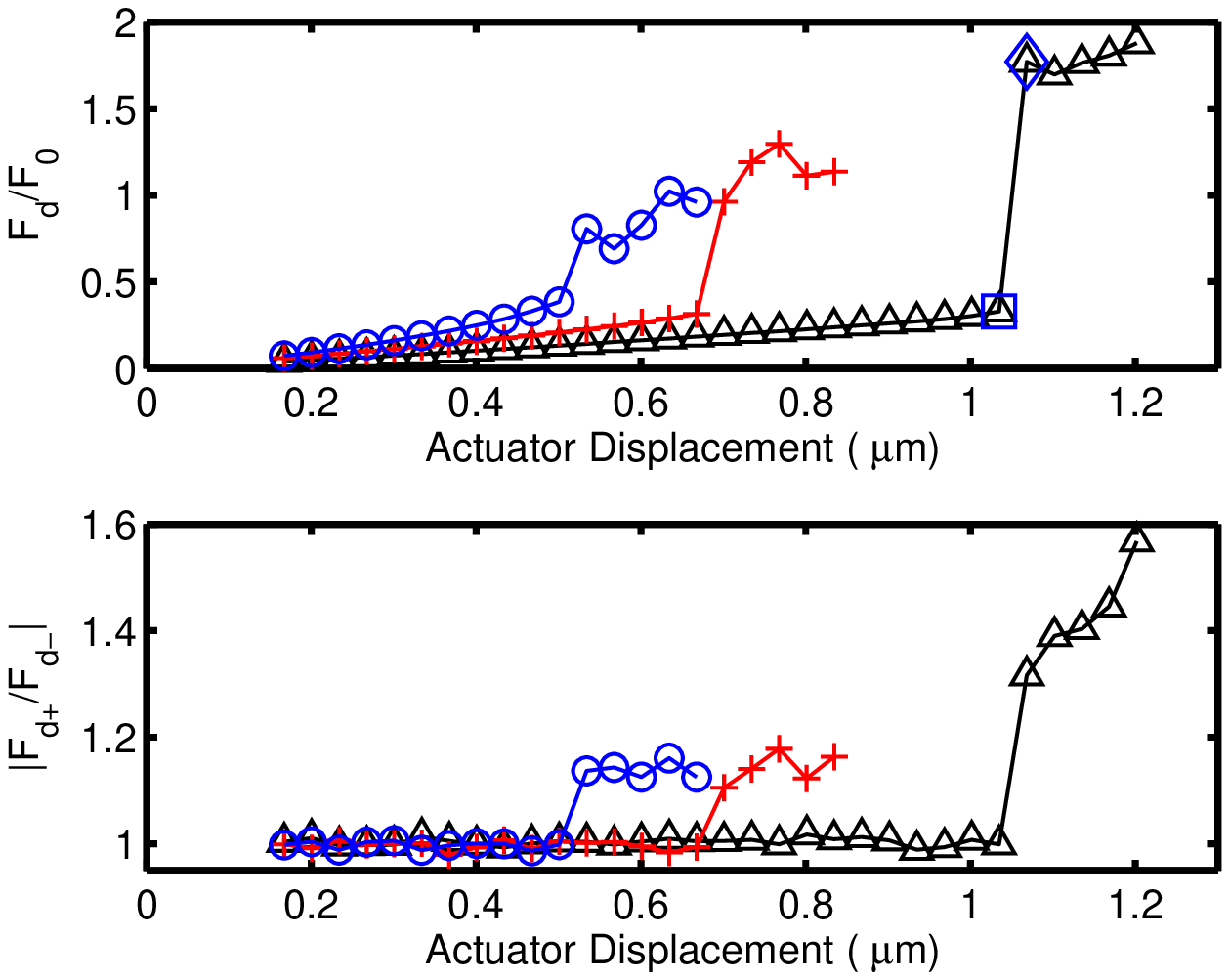}
\includegraphics[width=.2\textwidth]{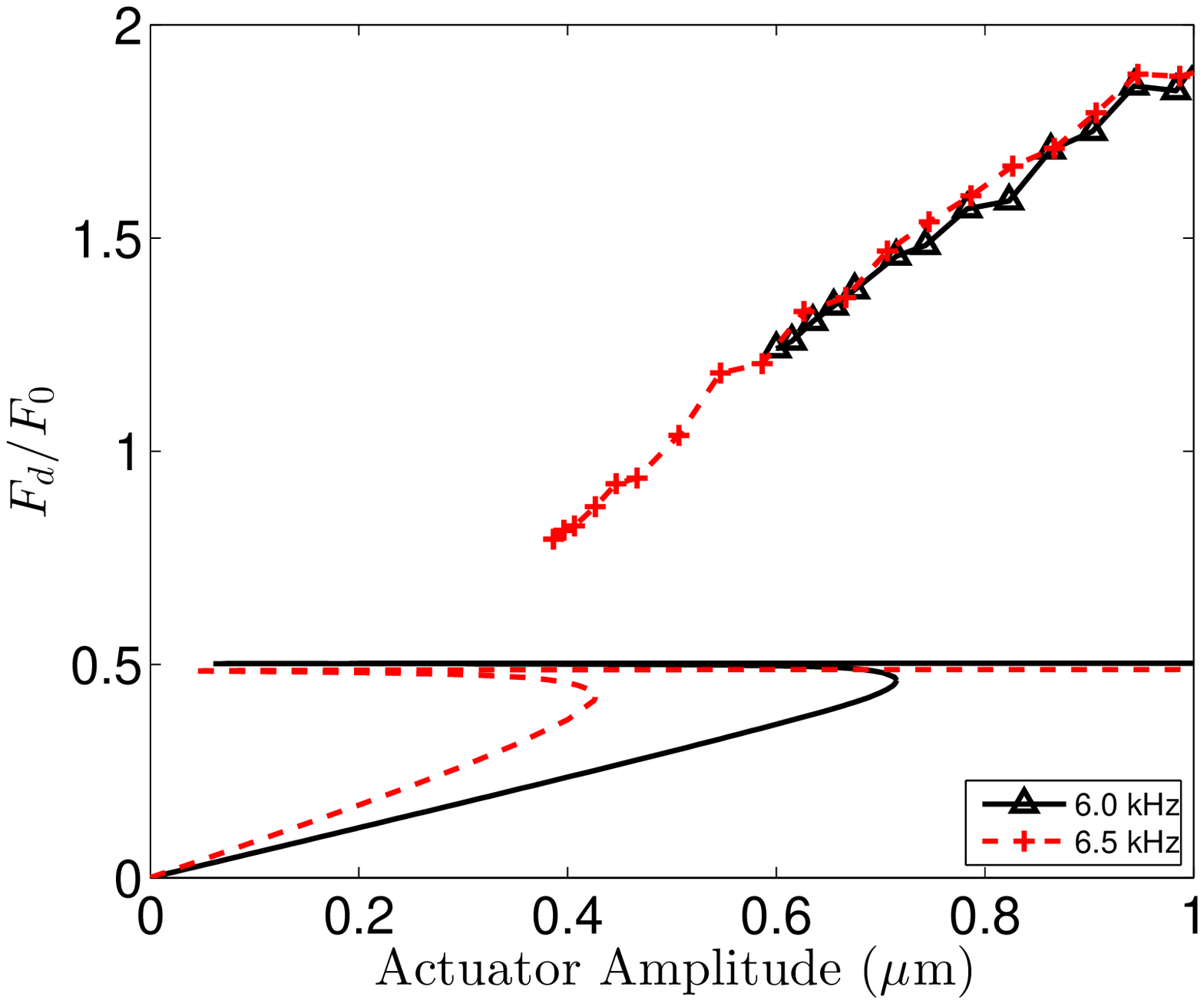}
\caption{\label{f:bifurc_diagram}
	(Left panel-top) Dynamic force amplitude at driving frequencies of $f_d =$ 6 kHz (black triangles), 6.5 kHz (red crosses), and 6.8 kHz (blue circles). 
	The blue square corresponds to the force-time history before the onset of chaos (top left plot of Fig. \ref{f:force_time}), while the blue diamond corresponds to the force-time history immediately following the onset of chaos (bottom left plot of Fig. \ref{f:force_time}). 
	(Left panel-bottom) Asymmetry of the force-time response, as the actuator displacement $\alpha$ is increased.
	The onset of asymmetry (as $|F_{d+}/F_{d-}|$ becomes $>1$) corresponds to the onset of instability for each of the three cases of driving frequency.
	(Right panel) Numerically obtained dynamic force amplitude at driving frequencies of $f_d =$ 6 kHz (black solid line with triangles) and 6.5 kHz (red dashed line with crosses). In each case, the line without markers corresponds to the numerically exact periodic solutions, while the line with markers corresponds to the chaotic branch.}
\end{figure}

	We estimate the average dynamic force amplitude as $\sqrt{2}$ times the root mean square (RMS) variation of the time interval
highlighted in blue (from $t=90$ ms to $t=122.8$ ms) of the force-time history. 
	In the top panel of Fig. \ref{f:bifurc_diagram}, we plot this dynamic force amplitude (normalized against the static load, $F_0$) as a function of $\alpha$. 
	We plot this response for driving frequencies of $f_{d} = 6$, $6.5$, and $6.8$ kHz. 

	For each driving frequency (6 kHz [black triangles], 6.5 kHz [red crosses], and 6.8 kHz [blue circles]), after $\alpha$ crosses a certain threshold, the system jumps from a low amplitude stable periodic response to a high amplitude chaotic response (top and bottom left panels of Fig. \ref{f:force_time}). 
	This transition is also characterized by a drastic change in the frequency content of the response, as shown in the right panel of Fig. \ref{f:force_time}, from that of a single frequency response corresponding to the driving frequency (black - before instability) to a broadband frequency response (blue - after instability). 
	The change in the frequency content of the response implies transfer of the applied energy to frequency components that belong to the propagating bands. 
	This leads to energy propagation in forbidden band gaps. 
	A similar phenomenon, although of a different origin, has been observed experimentally on the chain of coupled pendula \cite{geniet2002}.

	The trend is similar to that of the numerical results shown in the right panel of Fig \ref{f:bifurc_diagram} where an increase in $f_d$ corresponds to a decrease in $\alpha_c^{(1)}$. 
	Quantitative differences between experiments and numerics can be explained by the observed differences between theoretical and experimental spectra. 
	Driving the chain at $6$ kHz numerically, where the lower cutoff frequency is at $6.96$ kHz, is not equivalent to driving it at $6$ kHz experimentally, where the lower cutoff frequency found to be around $7.83$ kHz or $8.5$ kHz before and after the jump to the chaotic branch respectively. 
	Other simplifications in the modeling of the setup, such as the approximation of dissipation with a linear on-site dissipative term and the consideration of a purely 1D geometry of the chain, may also contribute to the qualitative agreement (but small quantitative disagreement) between numerics and experiments. 

	Another remarkable characteristic that one can observe in the right panel of Fig \ref{f:bifurc_diagram}, is the fact that the chaotic branch seems to be very similar for both frequencies. 
	In numerics, we observed that is quite generic; the chaotic branch appears very similar for sufficiently large actuator amplitudes, independent of the driving frequency. 
	We have also observed that the higher the driving frequency is, the smaller the $\alpha_c^{(2)}$ and the larger the domain of existence of the chaotic branch. 
	Looking at the top-left panel of Fig \ref{f:bifurc_diagram}, one can observe a similar trend in the experiments. 
	Our intuition is that the chaotic branch is less sensitive to the driving frequency and depends mostly on gap openings between the beads. 
	This intuition is supported by the observed asymmetry of the force-time response which is shown in the bottom-left panel of Fig \ref{f:bifurc_diagram}. 
	The asymmetry of the force-time response is indicative of dynamic loss of contact of particles in the granular crystal. 
	Using the same portion of the force-time used to calculate the PSD, we separated the dynamic force signal into its positive and negative values, calculated the RMS of both ($F_{d+}$ and $F_{d-}$, respectively), and found the absolute value of the ratio between them, to give an effective measure of asymmetry ($|F_{d+}/F_{d-}|$). 
	For the time interval where the forces are symmetric, 
we expect this ratio to be close to $1$, indicating similarity between the positive and negative envelope of the signal. 
	However, we see that after instability occurs and system is driven to its chaotic branch of solutions, the ratio suddenly jumps from a value of nearly 1 to a substantially larger value. 
	This phenomenon is seen in all three driving frequencies, and the onset of asymmetry occurs at the same actuator displacements as those of the system's transitions to chaos. 

\begin{figure}[h!]
\includegraphics[width=.4\textwidth]{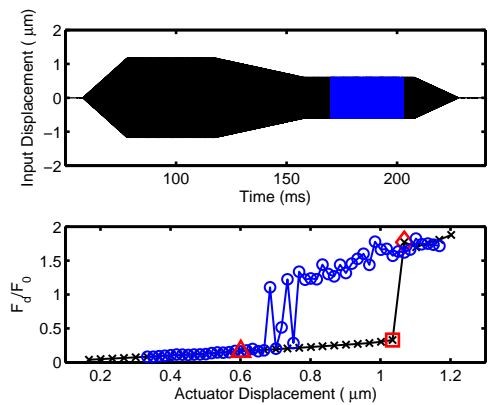}
\caption{\label{f:hysteretic_loop} 
	(Top) The applied actuator driving amplitude (displacement) for the hysteresis loop at the trial marked by the red triangle of the bottom figure. 
	The initial amplitude level is $1.17$ $\mu$m for all experiments. 
	The blue is the region over which the Power Spectral Density (PSD) is calculated.
	(Bottom) Experimentally characterized hysteresis loop as a function of actuator driving amplitude (displacement), at a driving frequency of 6 kHz. 
	Black x markers denote following the stable periodic solution into chaos, while blue circles denote following the return from a chaotic response back down to the low amplitude periodic response. 
	The red square corresponds to the blue square of Fig. \ref{f:bifurc_diagram}, while the red diamond corresponds to the blue diamond of Fig. \ref{f:bifurc_diagram}.}
\end{figure}

	Finally, after measuring the bifurcation of the system response for the $f_d = 6$ kHz case, we characterize the return part of the hysteresis loop of our granular crystal response, by exciting the crystal with a new signal of the same frequency and a duration of $170$ ms.
	The amplitude of the signal is first linearly ramped up for $20$ ms to a voltage high enough to induce chaos. 
	This amplitude is held constant for $40$~ms, and then is linearly ramped down (over $40$ ms) to a lower secondary amplitude (which is gradually decreased over each sequential run) that continues for $50$ ms. 
	Finally, the last $20$ ms are linearly ramped down to zero amplitude. 
	These linear ramps allow us to track the low amplitude stable periodic branch until the point of its disappearance. After this, we can follow the chaotic solution down to lower amplitudes, until the system response drops back to the low amplitude branch of periodic solutions. 
	The average dynamic force is calculated and measured as described previously, except here the middle of the secondary constant amplitude level is used for the PSD (from $t=170$ ms to $t=202.8$ ms, as shown in the top panel of Fig. \ref{f:bifurcation_diagram}). 
	The experimentally measured hysteresis loop is shown in the bottom panel of Fig. \ref{f:hysteretic_loop} for a driving frequency of 6 kHz. 
	As we track the chaotic branch downwards, there is an intermediate region over which the solutions are observed to alternate between chaotic and periodic responses ($\sim0.7$ and $\sim0.8$ $\mu$m), before settling back to the stable low amplitude periodic branch. 
	These jumps ``back up'' to chaos may simply be a sign of a transient chaotic regime. 
	By keeping the driving at  that secondary amplitude longer, the system may have eventually fallen back to the periodic response instead. 
	It could also be attributed to small differences in the geometry of the chain from measurement to measurement induced by 
the 
reassembling of the granular crystal due to the high amplitude dynamic loading.
	
%

{\it Conclusions and Future Challenges.} \label{S:______Conclusion}
	In this work, we have provided a prototypical case example where the interplay of the damped-driven dynamics (at least partially extrinsic to the chain) and the periodic nonlinear dynamics (chiefly intrinsic to the chain) give rise to saddle-node bifurcations of periodic solutions 
and turning points of ordered dynamics. 
	These terminations can be controlled/manipulated via relevant parameters (such as the frequency of the drive).
	Past these points, the response of the system is chaotic, yet it can also be chaotic well before that, creating regimes of multi-stability and associated separatrices between time-periodic and genuinely chaotic attractors.
	This study provides a gateway for the exploration and controllable utilization of the capability of the system to exhibit large amplitude 
chaotic dynamics, as well as (and even concurrently with) periodic in time
dynamics. 

	Nevertheless, numerous interesting questions remain unexplored.
	The mechanism leading to the emergence of chaotic dynamics is a paradigm of an issue that certainly warrants further investigation. 
	An avenue that we haven't explored in detail herein (but will present in detail in a future investigation) involves the emergence of quasi-periodic solutions from the critical (Hopf) point of $\alpha \approx 1.7 \times 10^{-7}$ m. 
	It is plausible -yet presently unclear- that sequences of global bifurcations destroy these tori, eventually leading to chaotic dynamics.
	It is also conceivable that the stable and unstable manifolds of the ``dash-dot'' (upper) branch of periodic solutions are involved in the emergence, as the drive amplitude is increased (and disappearance as it is decreased), of the apparent chaotic attractors. 
	This intriguing, yet technically challenging theme will be deferred to future investigations.

{\it Acknowledgements.} CD acknowledges support
from the National Science Foundation, grant number CMMI-844540 (CAREER) and NSF
1138702. PGK acknowledges support from the US National 
Science Foundation
under grant CMMI-1000337, the US Air Force under grant FA9550-12-1-0332, 
the Alexander von Humboldt Foundation, as well as the Alexander S. Onassis 
Public Benefit Foundation. IGK acknowledges support from the
 US Air Force under grant FA9550-12-1-0332.
GT acknowledges support from the Alexander S. Onassis 
Public Benefit Foundation.

\end{document}